\documentclass{svjour3}                     
\smartqed  
\usepackage{graphicx}
\usepackage{microtype}
\usepackage[utf8x]{inputenc}
\usepackage[english]{babel}
\usepackage[babel]{csquotes}
\usepackage{authblk}
\usepackage{mathrsfs}
\usepackage{indentfirst}
\usepackage{amssymb,amsmath}
\usepackage{array}
\usepackage{epsfig}
\usepackage{latexsym}
\usepackage{array}
\usepackage{fancyhdr}
\usepackage{pstricks}
\usepackage{bm}
\usepackage{slashed}
\usepackage{color}
\usepackage{booktabs}
\usepackage{braket}

\newcommand{\beq}{\begin{equation}}
\newcommand{\eeq}{\end{equation}}
\newcommand{\eq}{Eq.~}

\newcommand{\fig}{Fig.~}

\begin{document}

\title{Polarization and dilepton angular distribution in pion-nucleon collisions
}

\titlerunning{Polarization and 
  dilepton angular distribution in $\pi+N$ collisions}        

\author{Mikl\'os Z\'et\'enyi   \and Enrico Speranza \and Bengt Friman
}


\institute{Mikl\'os Z\'et\'enyi \at
  Wigner Research Center for Physics, H-1121 Budapest, Hungary \\
              \email{zetenyi.miklos@wigner.mta.hu}
           \and Enrico Speranza
            \at
            Institut f\"ur Theoretische Physik, Goethe-Universit{\"a}t, D-60438 Frankfurt am Main, Germany \\
            \and Bengt Friman
            \at
            GSI Helmholtzzentrum f\"ur Schwerionenforschung GmbH, D-64291 Darmstadt, Germany
}

\date{Received: date / Accepted: date}

\maketitle

\begin{abstract}
We study hadronic polarization and the related anisotropy of the dilepton angular distribution for the
reaction $\pi N \to Ne^+e^-$. We employ consistent effective interactions for baryon resonances up to spin-5/2
to compute their contribution to the anisotropy coefficient. We show that the spin and parity of the
intermediate baryon resonance is reflected in the angular dependence of the anisotropy coefficient. We
present results for the anisotropy coefficient including the $N(1520)$ and $N(1440)$ resonances, which are
essential at the collision energy of the recent data obtained by the HADES collaboration on this reaction.
We conclude that the anisotropy coefficient provides useful constraints for unraveling the resonance
contributions to this process.
\end{abstract}

\section{Introduction}
Dilepton production in hadronic reactions provides information on the 
electromagnetic properties of hadrons.  Dileptons are produced in a variety of different elementary 
processes. Independently of the specific reaction, they originate from the decay of virtual photons. Multiply differential cross sections for dilepton production can provide 
information that may help to disentangle the production channels. 

The HADES collaboration has recently studied dilepton production in pion-induced reactions \cite{HADES}. 
One of the most important contributions to this reaction originates from processes with intermediate $s$-channel baryon resonances, $\pi N \rightarrow R \rightarrow Ne^+e^-$.
In this presentation we explore the baryonic contributions to the reaction
$\pi N \rightarrow Ne^+e^-$ in terms of an effective Lagrangian model 
at the center-of-momentum (CM) energy of the HADES experiment. In 
particular, we study the angular distribution of the produced dileptons, which reflects
the polarization state of the decaying virtual photon, and, ultimately, carries information about the process
in which the virtual photon was created. In the present case, the polarization data provides constraints on the quantum numbers of the intermediate 
baryon resonances. This contribution is based on our previous publication \cite{PLB}, where the interested
reader can find further details.

\begin{figure}[tb]
   \begin{center}
   \scalebox{0.2}{\includegraphics{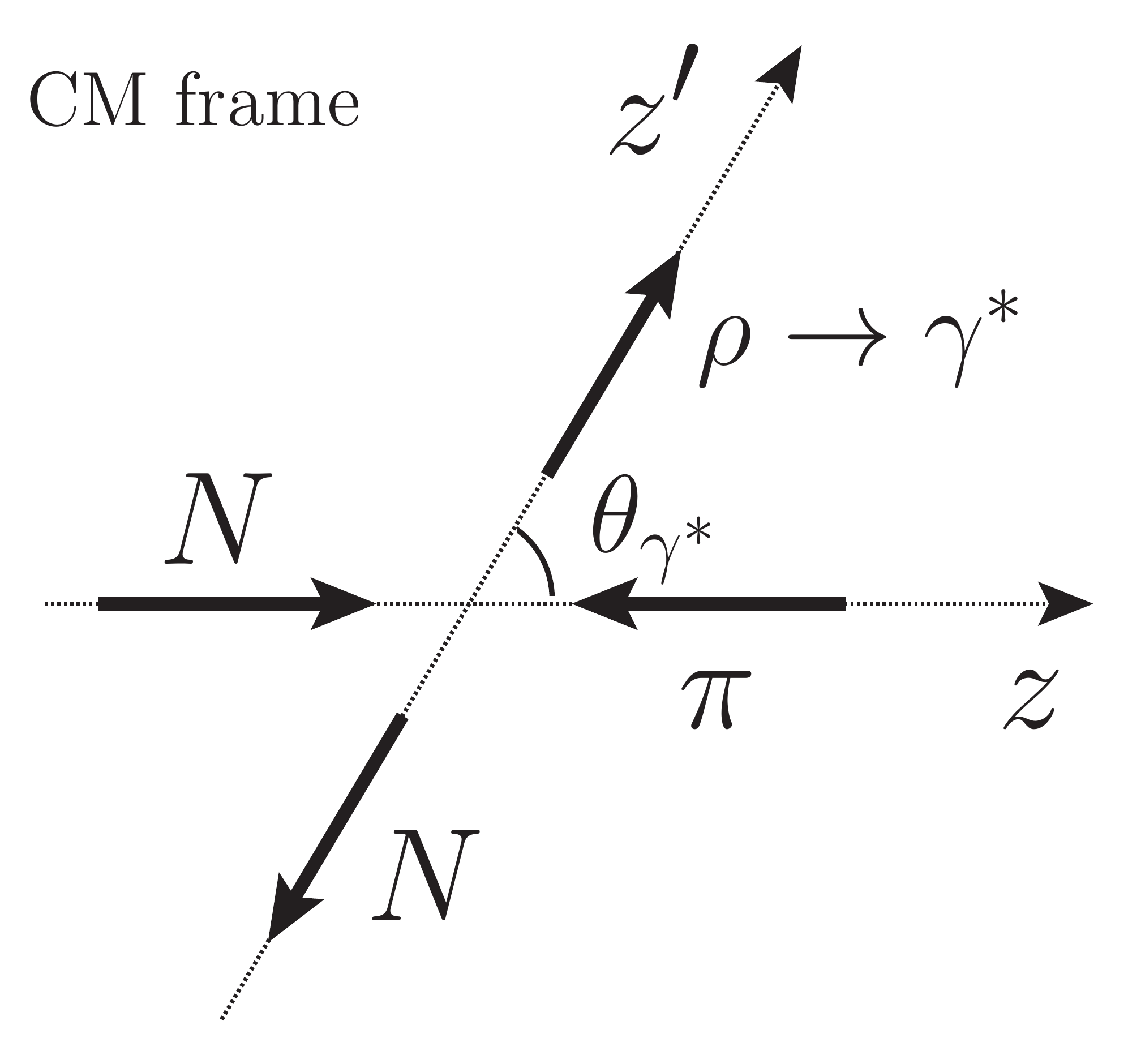}}\hfil
   \scalebox{0.2}{\includegraphics{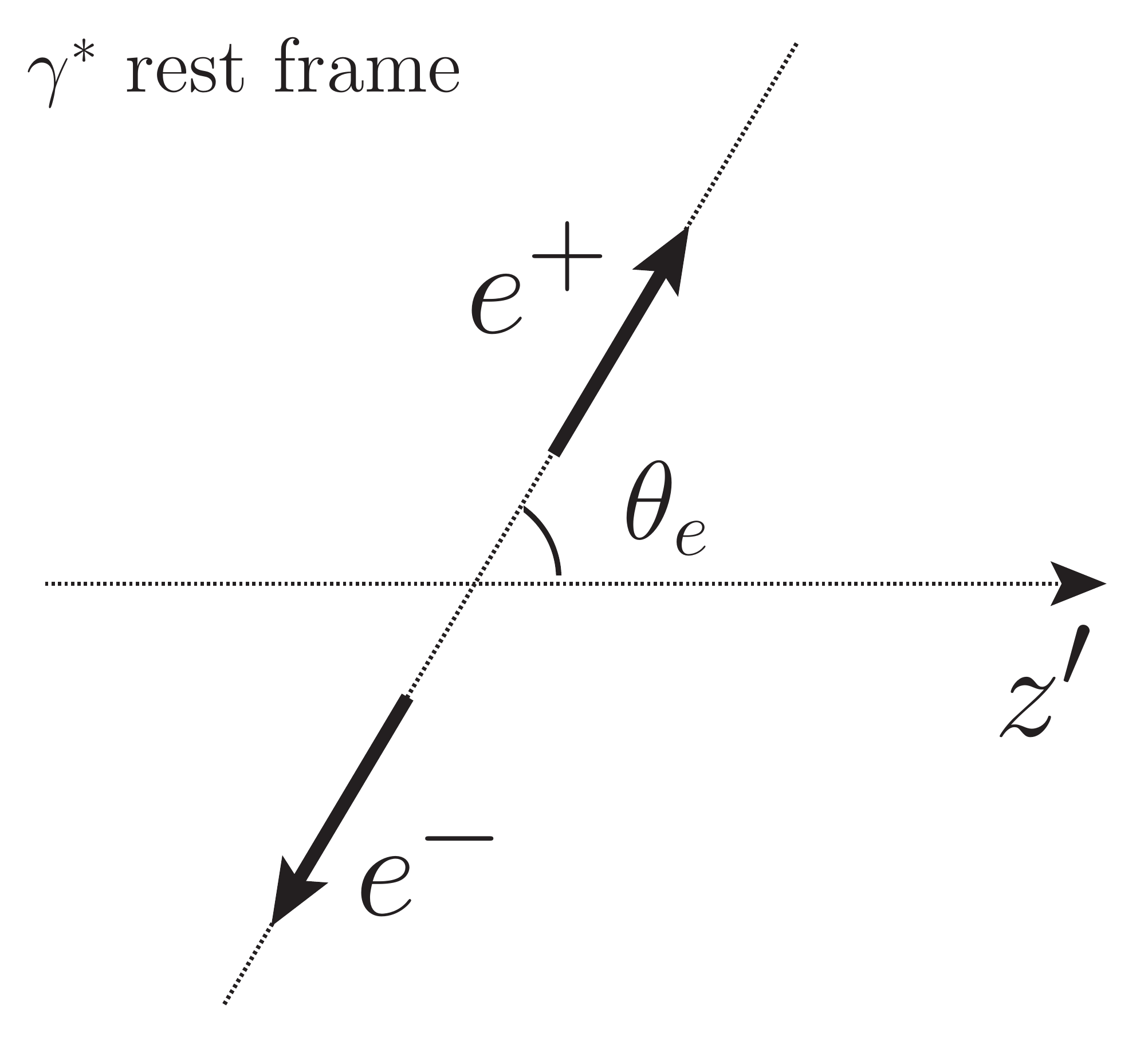}}
   \end{center}
   \caption{\label{fig:angles} Definition of the angles in the CM frame and the virtual photon rest frame.}
\end{figure}

The general expression for the triple differential cross section of the process $\pi N \to Ne^+e^-$ is
\begin{align}
  \label{eq:angdist}
  \frac{d\sigma}{dM d\cos\theta_{\gamma^*} d\cos_e}& \propto  
      \Sigma_\bot (1+\cos^2 \theta_e) + \Sigma_\parallel  (1-\cos^2 \theta_e) \nonumber \\
  & \propto A(1+\lambda_\theta(\theta_{\gamma^*},M)\cos^2 \theta_e),
\end{align}
where $M$ is the mass of the virtual photon (the dilepton invariant mass), $\theta_{\gamma^*}$ is the virtual
photon polar angle in the CM frame, and $\theta_e$ is the polar angle of one
of the two leptons in the rest frame of the photon as measured relative to the momentum of the virtual photon in the CM frame. This choice is called the helicity frame (see Fig.~\ref{fig:angles}).
In the second line we defined the anisotropy coefficient 
\begin{equation}
     \lambda_\theta(\theta_{\gamma^*},M) = \frac{\Sigma_\bot - \Sigma_\parallel}{\Sigma_\bot + \Sigma_\parallel},
\end{equation}
which contains information on the polarization of the virtual
photon and hence on the reaction mechanism.

The emergence of a dilepton anisotropy in the process 
$\pi N \rightarrow R \rightarrow Ne^+e^-$
can be understood as follows.
The initial state, which in the CM frame contains a pion with momentum $\mathbf{p}$
and a nucleon with momentum $-\mathbf{p}$, can be expanded
in terms of eigenstates of orbital angular momentum
\begin{equation}
  \label{eq:orbital}
  \Ket{\pi(\mathbf{p});N(-\mathbf{p})} \propto 
  \sum_{lm} Y^*_{lm}(\theta,\phi) \Ket{lm},
\end{equation}
where $\theta$ and $\phi$ specify the direction of $\mathbf{p}$ with
respect to the quantization axis. We choose the quantization axis $z$
parallel to the momentum of the incident pion, implying that $\theta = 0$. In this case the $z$-component of the
orbital angular momentum vanishes in the initial state. Hence, the projection 
of the total spin of the intermediate
baryon resonance on the beam axis is the same as the $z$-component of the nucleon
spin. This means that only the $J_z=+1/2$ and $-1/2$ states of the resonance are 
populated. 

As a result, in case of an unpolarized nucleon target, spin-$1/2$ 
intermediate resonances are unpolarized, and consequently there is no 
preferred direction in the CM frame. Accordingly, in this case all
observables are independent of the scattering angle, i.e., the
angle $\theta_{\gamma^*}$ of the virtual photon in the CM frame. 
On the other hand, intermediate resonances of spin$\ge 3/2$ have a 
nontrivial polarization, implying an angular anisotropy 
in the CM frame. Consequently, in this case, observables show a nontrivial dependence
on the scattering angle $\theta_{\gamma^*}$, 
which reflects the quantum numbers of the resonance.

\section{\label{sec:xsec} Cross section and anisotropy coefficient}
The matrix element of the process $\pi N \rightarrow
Ne^+e^-$ can be decomposed as a product of a leptonic part, describing the decay of the virtual photon, 
and a hadronic part, containing the rest of the process including the production of the virtual photon.
Consequently, the square of the matrix element can be written in the form
\begin{equation}\label{eq:msquared}
  \sum_\text{pol} \left\vert\mathcal{M}\right\vert^2 = \frac{e^2}{k^4}
  W_{\mu\nu}l^{\mu\nu},
\end{equation}
where the lepton tensor, $l^{\mu\nu}$, is determined by the leptonic matrix element. Its explicit form
is easily obtained from quantum electrodynamics, as
\begin{equation}
  \label{eq:lmunu}
  l^{\mu\nu} = 4\left[k_1^{\mu}k_2^{\nu} + k_1^{\nu}k_2^{\mu} -
  (k_1\cdot k_2+m_e^2)g^{\mu\nu} \right],
\end{equation}
where 
$k_1$ and $k_2$ are the momenta of the electron and positron, respectively, and $m_e$ is the electron mass.
The quantity
\begin{equation}
  W_{\mu\nu} = \sum_\text{pol}
  \mathcal{M}_{\mu}^{\text{had}}{\mathcal{M}_{\nu}^{\text{had}}}^{*}
\end{equation}
is the hadronic tensor, expressed in terms of the hadronic matrix element, $\mathcal{M}_{\mu}^{\text{had}}$.

It is convenient to introduce the polarization density matrix formalism \cite{Choi:1989yf}. The hadronic (or production) density
matrix is defined as 
\begin{equation}
  \rho^{\text{had}}_{\lambda,\lambda^{\prime}} = \frac{e^2}{k^4} 
  \epsilon^{\mu}(k,\lambda) W_{\mu\nu} \epsilon^{\nu}(k,\lambda^{\prime})^*,
\end{equation}
and it represents the polarization state of the virtual photon produced in the hadronic part of the reaction.
The leptonic (or decay) density matrix is defined as
\begin{equation}
  \rho^{\text{lep}}_{\lambda^{\prime},\lambda} = 
  \epsilon^{\mu}(k,\lambda^{\prime}) l_{\mu\nu} 
  \epsilon^{\nu}(k,\lambda)^*,
\end{equation}
and it projects the above state of the virtual photon on the final state containing the lepton pair.
In the above, $\epsilon^{\mu}(k,\lambda)$ denotes the polarization
vector of the virtual photon of momentum $k$ and helicity $\lambda$. The latter can 
take on values $\pm$1 and 0, since virtual photons can also be longitudinally polarized.
In terms of the polarization density matrices, the square of the matrix element - which contains all
the information about angular distributions - can be written as
\begin{equation}
  \label{eq:M2}
  \sum_\text{pol} \left\vert\mathcal{M}\right\vert^2 = 
    \sum_{\lambda,\lambda^{\prime}}
  \rho^{\text{had}}_{\lambda,\lambda^{\prime}} 
    \rho^{\text{lep}}_{\lambda^{\prime},\lambda}.
\end{equation}

Using the explicit form of the lepton tensor, Eq.~\ref{eq:lmunu}, and neglecting the electron mass,
the leptonic density matrix is obtained as
\begin{equation}
  \label{eq:rho_lep}
  \rho^{\text{lep}}_{\lambda^{\prime},\lambda} = 4|{\bf k}_1|^2 \left(
  \begin{array}{ccc}
    1+\cos^2\theta_e & -\sqrt{2}\cos\theta_e\sin\theta_e e^{-i\phi_e} & 
      \sin^2\theta_e e^{-2i\phi_e} \\
    -\sqrt{2}\cos\theta_e\sin\theta_e e^{i\phi_e} & 2(1-\cos^2\theta_e) &
      \sqrt{2}\cos\theta_e\sin\theta_e e^{-i\phi_e} \\
    \sin^2\theta_e e^{2i\phi_e} & 
      \sqrt{2}\cos\theta_e\sin\theta_e e^{i\phi_e} & 1+\cos^2\theta_e 
  \end{array}
  \right),
\end{equation}
where ${\bf k}_1$ is the three-momentum of one of the two leptons in the virtual photon rest frame. 
The angular dependence of the squared matrix element is obtained by 
combining Eqs.\ \eqref{eq:M2} and \eqref{eq:rho_lep},
\begin{align}
	\label{eq:msq_spin-dens}
  \sum_\text{pol} \left\vert\mathcal{M}\right\vert^2 & \propto
  (1+\cos^2\theta_e )(\rho^{\text{had}}_{-1,-1} + \rho^{\text{had}}_{1,1})
  + 2(1-\cos^2\theta_e )\rho^{\text{had}}_{0,0} \nonumber\\
    & + \textrm{terms vanishing upon integration over } \phi_e.
\end{align}
Here we suppressed the explicit dependence of $\rho^{\text{had}}_{\lambda,\lambda^{\prime}}$ on $M$ and $\theta_{\gamma^*}$.
By comparing Eqs.~(\ref{eq:angdist}) and (\ref{eq:msq_spin-dens}), we can identify
the anisotropy coefficient
\begin{equation}
  \label{eq:bcoeff}
  \lambda_\theta  = \frac{\rho^{\text{had}}_{-1,-1}+\rho^{\text{had}}_{1,1}-2\rho^{\text{had}}_{0,0}}
  {\rho^{\text{had}}_{-1,-1}+\rho^{\text{had}}_{1,1}+2\rho^{\text{had}}_{0,0}}.
\end{equation}

\section{\label{sec:model} The effective Lagrangian model}
In this section we specify the terms of the effective Lagrangian describing the interactions of all hadrons
playing a role in the process. From this Lagrangian, transition matrix elements and also the hadronic density
matrix can be calculated via standard Feynman diagram techniques.

We assume that baryons couple to the electromagnetic field via an
intermediate $\rho^0$ meson according to the vector meson dominance
model. For the $\rho^0$-photon coupling we use the gauge invariant vector meson dominance 
model Lagrangian~\cite{Kroll:1967it}
\begin{equation}
  \label{eq:VMD}
  \mathcal{L}_{\rho\gamma} = - \frac{e}{2g_{\rho}} F^{\mu\nu}
  \rho^0_{\mu\nu},
\end{equation}
where $F^{\mu\nu} = \partial_{\mu}A_{\nu} - \partial_{\nu}A_{\mu}$ is
the electromagnetic field strength tensor and $\rho^0_{\mu\nu} =
\partial_{\mu}\rho^0_{\nu} - \partial_{\nu}\rho^0_{\mu}$.

We include baryon resonances up to spin-5/2. The interaction of spin-1/2 baryons with 
pions and $\rho$ mesons is described by the Lagrangian densities of 
Ref.~\cite{Zetenyi:2012hg},
\begin{eqnarray}
  \mathcal{L}_{R_{1/2}N\pi} & = & - \frac{g_{RN\pi}}{m_{\pi}} 
  \bar{\psi}_R \Gamma
  \gamma^{\mu}\vec{\tau}\psi_N \cdot \partial_{\mu}\vec{\pi}
  + \text{h.c.}, \label{eq:RNpi_1h} \\
   \mathcal{L}_{R_{1/2}N\rho} & = & \frac{g_{RN\rho}}{2m_{\rho}} 
  \bar{\psi}_{R}
  \vec{\tau} \sigma^{\mu\nu} \tilde{\Gamma} \psi_N \cdot \vec{\rho}_{\mu\nu} +
  \text{h.c.}. \label{eq:RNrho_1h}
\end{eqnarray}
Here, and also in the Lagrangians involving higher spin resonances given below,
$\Gamma = \gamma_5$ for $J^P = 1/2^+$, $3/2^-$ and $5/2^+$
resonances and $\Gamma = 1$ otherwise, and $\tilde{\Gamma} =
\gamma_5\Gamma$.

Higher spin fermions are represented by Rarita-Schwinger spinor fields in
effective Lagrangian models. We include spin-3/2 and spin-5/2 resonances in our model
and describe their interactions using the consistent interaction scheme developed by 
Vrancx et al.\ \cite{Vrancx:2011qv}. For details we refer to Ref.\ \cite{PLB}.

\section{\label{sec:results} Results}
We employ the model described above to compute the anisotropy
coefficient $\lambda_\theta$ of \eq\eqref{eq:bcoeff} for the reaction $\pi N \to N e^+
e^-$.  In the following we discuss the dependence of the anisotropy
coefficient on the polar angle of the virtual photon
$\theta_{\gamma^*}$. In all the calculations, the CM energy is set to
$\sqrt{s}=1.49$~GeV, corresponding to the HADES data.

\begin{figure}[bt]
	\begin{center}
		\scalebox{0.7}{\includegraphics{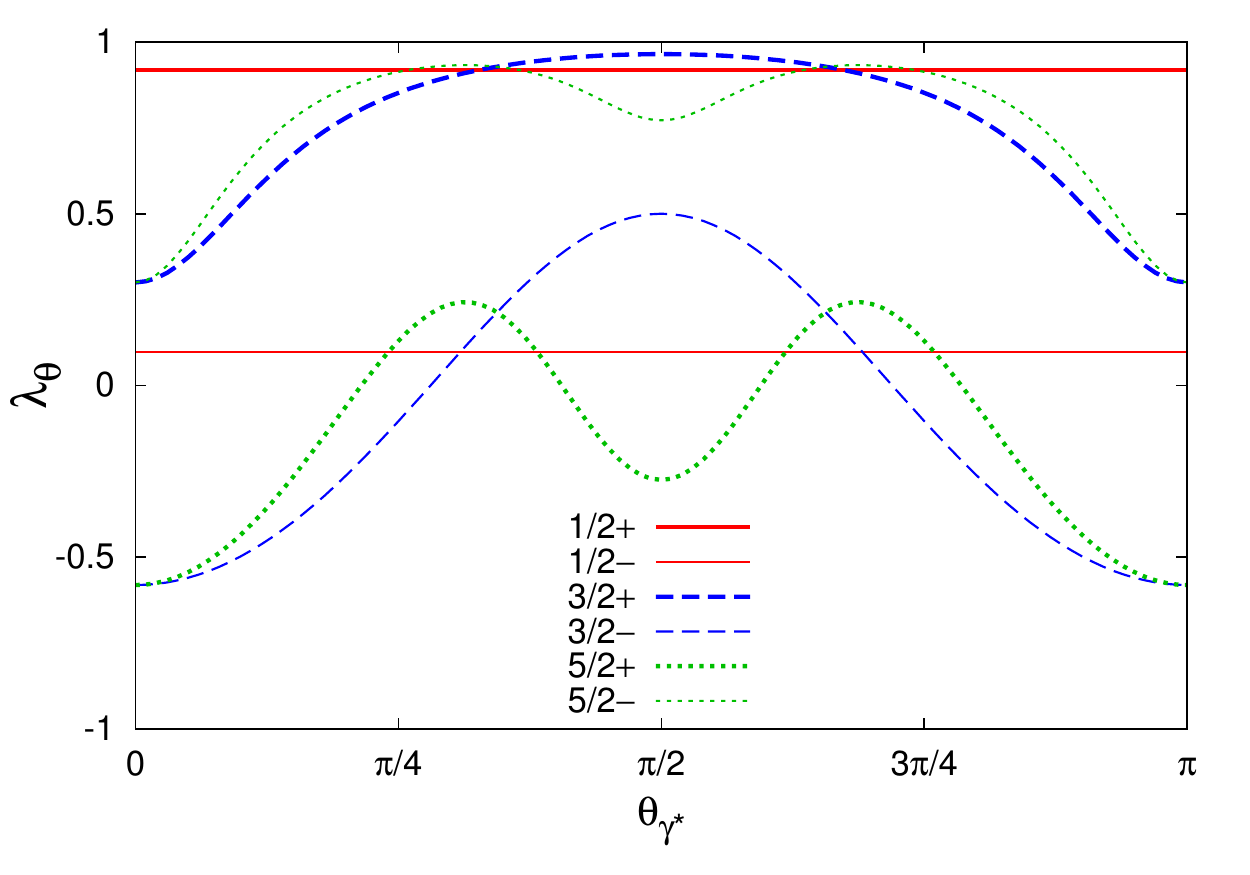}}
		\caption{\label{fig:schpole}The anisotropy coefficient $\lambda_\theta$ as a function of the virtual photon 
		  polar angle $\theta_{\gamma^*}$ for hypothetical 
			resonance states with different spins and parities in the $s$-channel at a dilepton mass
			$M=0.5$~GeV. The resonance masses coincide with 
			$\sqrt{s}=1.49$~GeV, the resonance widths are $\Gamma_R = 0.15$~GeV.}
	\end{center}
\end{figure}
We explore the relevance of the different spins and parities by
computing the anisotropy coefficient with a hypothetical resonance for each spin-parity
combination, with mass $m_R = 1.49$~GeV and width $\Gamma_R = 0.15$~GeV. 
The mass was chosen to coincide with the CM energy $\sqrt{s}$
used in our calculations, thus assuming that in the $s$-channel the resonance is on the mass shell.
The results of this calculation for dileptons of invariant mass
$M=0.5$~GeV are shown in Fig.~\ref{fig:schpole}. Here one can see that the spin and parity of the intermediate resonance is
reflected in a characteristic angular dependence of the anisotropy coefficient. 
In particular, in the spin-1/2
channels the $\lambda_\theta$ coefficient is independent of $\theta_{\gamma^*}$, in
accordance with the arguments given in the introduction.

\begin{figure}[bt]
	\begin{center}
		\scalebox{0.7}{\includegraphics{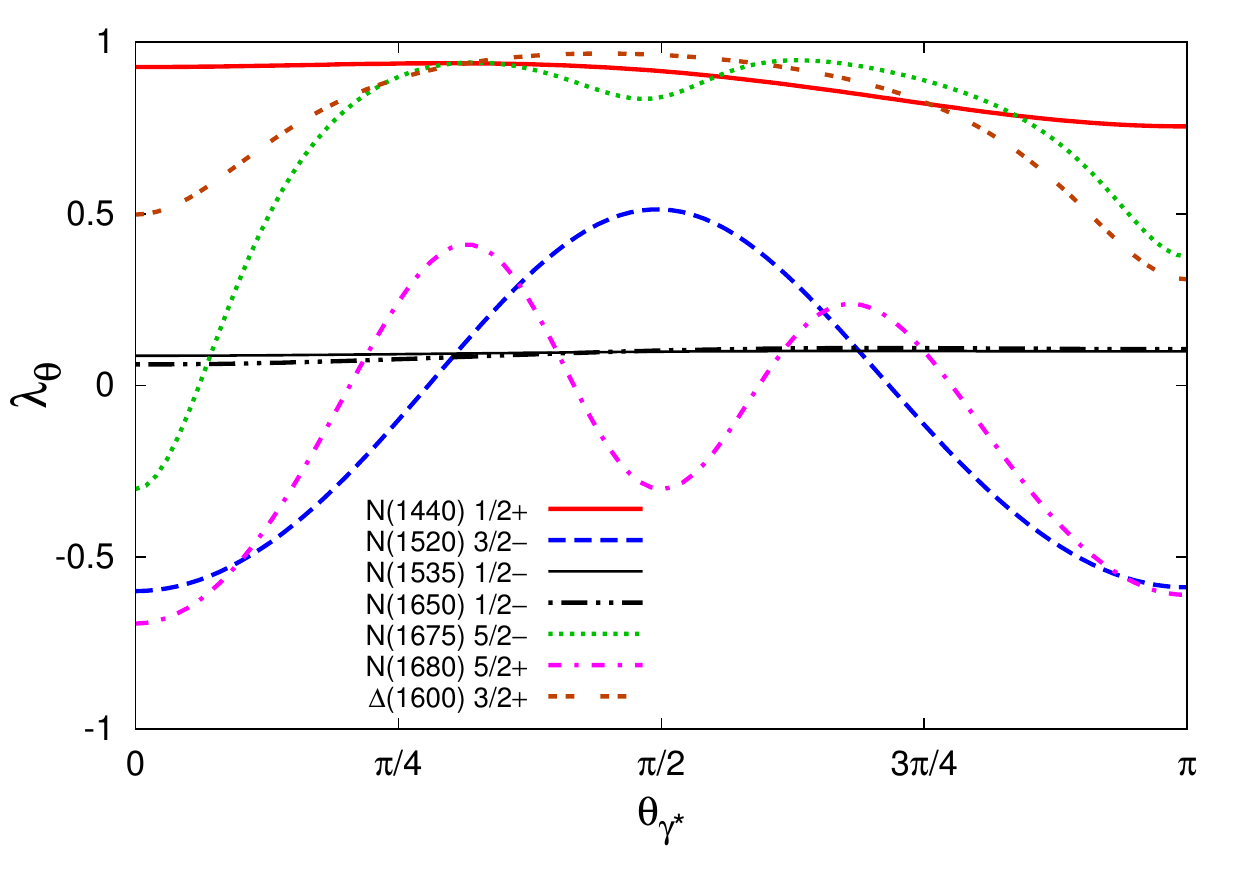}}
		\caption{\label{fig:sch_uch_physical} Contributions of resonances to the anisotropy coefficient 
		  $\lambda_\theta$ as a
			function of the virtual photon polar angle at a dilepton mass $M=0.5$~GeV
		including $s$- and $u$-channel diagrams. The CM energy is
			$\sqrt{s}=1.49$~GeV. For further details, see the text.}
	\end{center}
\end{figure}
In \fig\ref{fig:sch_uch_physical} we show the $\lambda_\theta$ coefficient obtained
from the $s$- and $u$-channel diagrams of the physical resonance
states that are relevant at the energy of the HADES experiment. 
Here, the characteristic shapes presented in \fig\ref{fig:schpole} are modified by the interference with the $u$-channel resonance contributions and the off-shellness of the $s$-channel contributions.

In order to identify the resonances that are important for 
the dilepton production process at the CM energy of the HADES experiment,
we compute the differential cross section $d\sigma/dM$ by integrating out the angles in the triple
differential cross section of Eq.~\ref{eq:angdist}. For this calculation, the coupling constants
$g_{RN\pi}$ and $g_{RN\rho}$ were determined from the widths of the
$R\rightarrow N\pi$ and $R\rightarrow N\rho\rightarrow N\pi\pi$ decays. The empirical values
for these partial widths were obtained as a product of the total width and the appropriate 
branching ratio as given by the Particle Data Group~\cite{PDG}.

We found that at $\sqrt{s} = 1.49$~GeV and a dilepton invariant 
mass of $M=0.5$~GeV, the two dominant contributions are due to the $N(1520)$
and $N(1440)$ resonances. 
\begin{figure}[bt]
	\begin{center}
		\scalebox{0.7}{\includegraphics{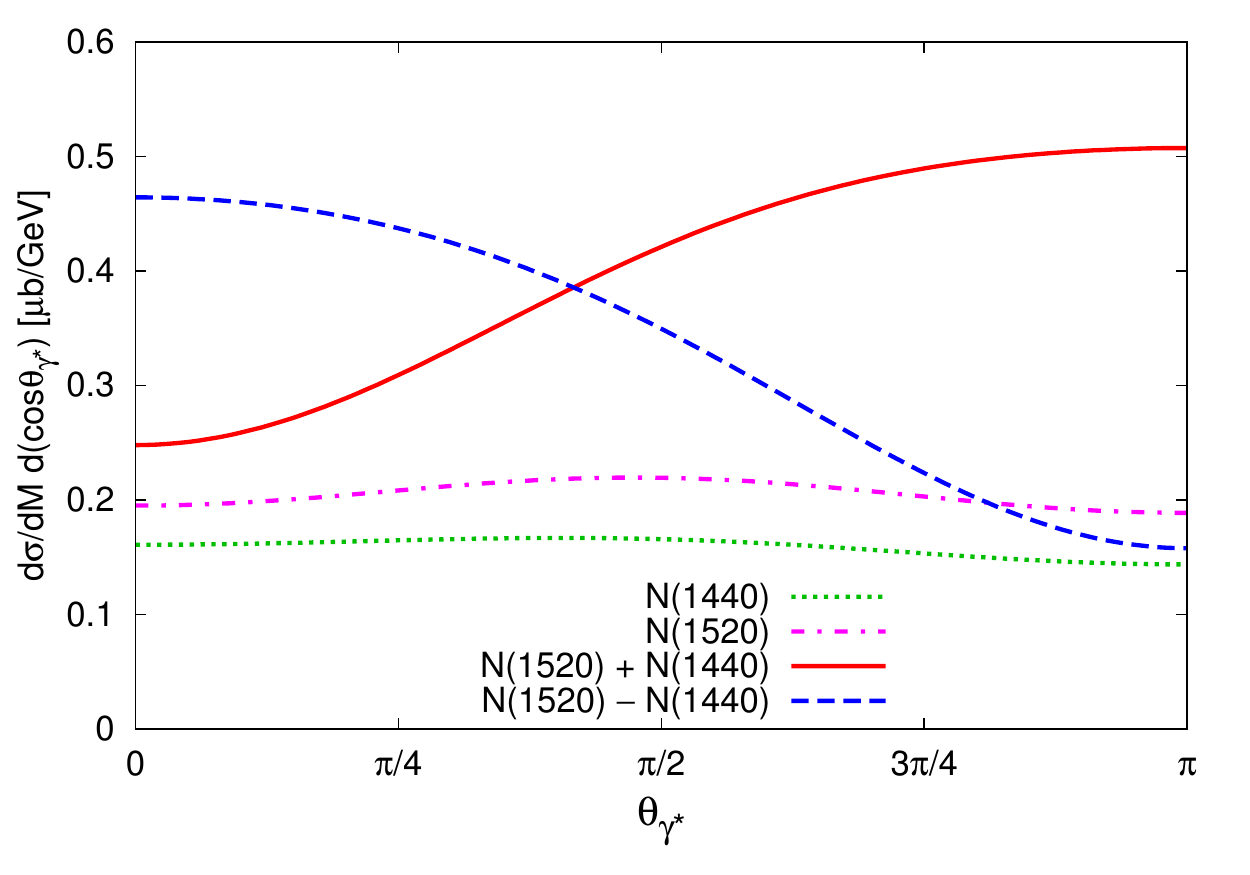}}
		\caption{\label{fig:sigma_phys_dominant} 
		  The contribution of the two dominant resonances, $N(1440)$ and $N(1520)$,
			to the differential cross section of dilepton production at $\sqrt{s}=1.49$~GeV
			CM energy and $M=0.5$~GeV dilepton mass. For the explanation of the various curves, see the text.}
	\end{center}
\end{figure}

The double-differential cross section, $d\sigma/dM d\cos\theta_{\gamma^*}$ obtained from
$s$- and $u$-channel diagrams with the dominant $N(1520)$ and $N(1440)$ resonances is shown in 
\fig\ref{fig:sigma_phys_dominant} as a function of the polar angle of the virtual photon, 
${\theta_{\gamma^*}}$. Here two of the curves correspond to the contributions of 
the two resonances without interference. In the other two, the interference terms are included, assuming either a positive or negative relative sign between the two resonance amplitudes.
From Fig.~\ref{fig:sigma_phys_dominant} it is clear that the relative phase has a strong 
influence on the shape of the $\theta_{\gamma^*}$ dependence
of the differential cross section. 

\begin{figure}[tb]
	\begin{center}
		\scalebox{0.7}{\includegraphics{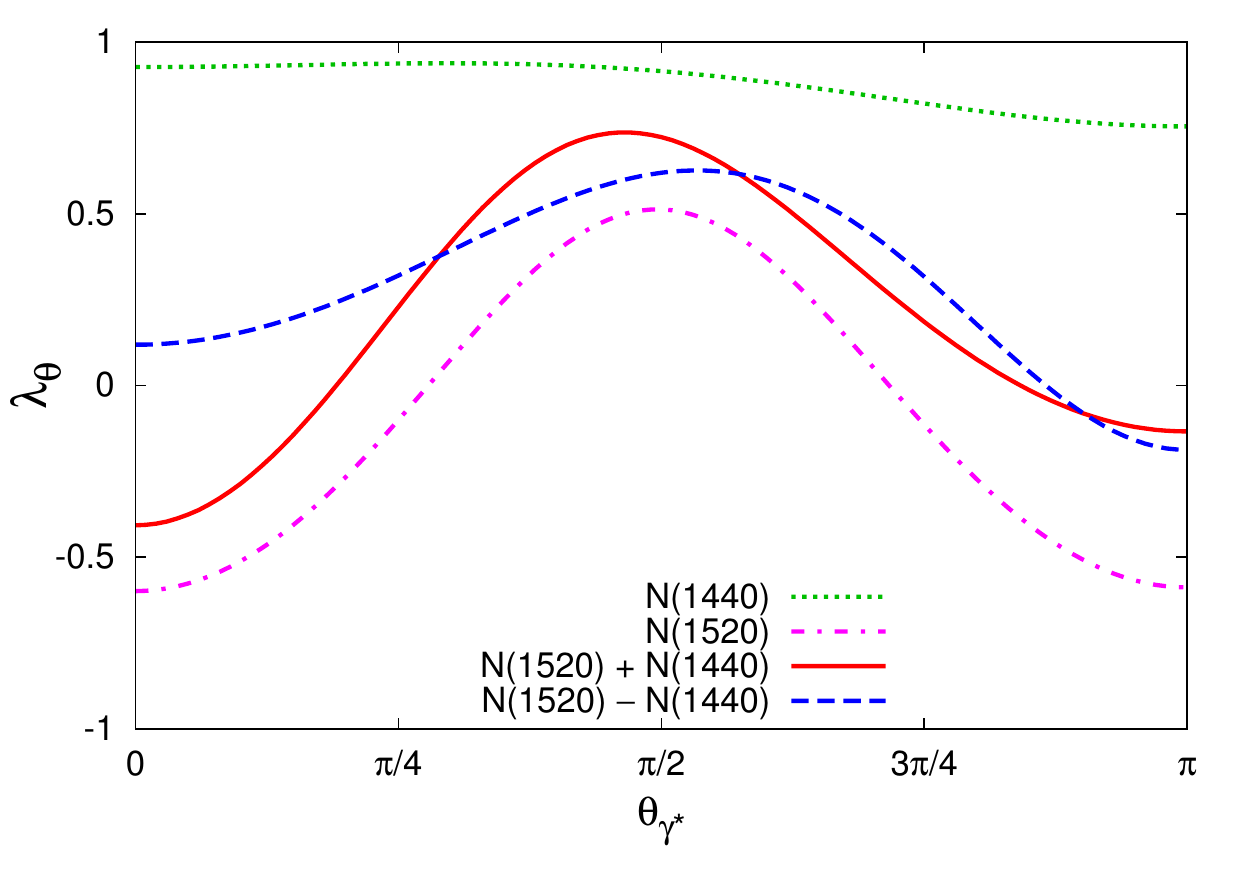}}
		\caption{\label{fig:B_phys_dominant} The contribution of the two dominant resonances, $N(1440)$ and $N(1520)$,
			to the anisotropy coefficient, $\lambda_\theta$ at $\sqrt{s}=1.49$~GeV CM energy 
			and $M=0.5$~GeV dilepton mass. 
			The various curves correspond to the same assumptions as in Fig.\ \ref{fig:sigma_phys_dominant}}.
	\end{center}
\end{figure}

In \fig\ref{fig:B_phys_dominant} we show the dominant contributions to the anisotropy coefficient $\lambda_\theta$ 
as a function of ${\theta_{\gamma^*}}$. As in Fig.~\ref{fig:sigma_phys_dominant}, we show
results for the two limiting  assumptions for the relative phase of the two resonance amplitudes. In both cases, the shape of the curve approximately follows that of  the
$N(1520)$ contribution, which implies that it is only weakly affected by the uncertainties of the $N(1440)$ parameters and the relative phase of the two amplitudes. The anisotropy parameter $\lambda_\theta$ has a maximum around $\theta_{\gamma^*}=\pi/2$, which means that virtual photons emitted 
perpendicular to the beam axis in the CM frame tend to be transversely polarized. On the other hand, 
virtual photons emitted along the beam direction are almost unpolarized or can even have some degree of longitudinal polarization.

\section{\label{sec:summary} Summary and outlook}

In this contribution we studied the angular distribution of dileptons originating 
from the process $\pi N \rightarrow Ne^+e^-$ and presented numerical results
for the anisotropy coefficient $\lambda_\theta$ based on the assumption that the process
is dominated by intermediate baryon resonances. Our results show that the shape of the anisotropy coefficient 
$\lambda_\theta$ as a function of the scattering angle can provide information on the spin-parity of the 
intermediate baryon resonance.

We estimated the coupling constants in our model based on the information on decay widths of baryon resonances
as given by the Particle Data Group. We find that at the HADES energy the cross section is dominated by 
the $N(1520)$ resonance and hence that the dilepton anisotropy reflects the $N(1520)$ with only a weak dependence on the model uncertainties.

The anisotropy coefficient can in principle be determined in 
experiments by the HADES collaboration at GSI, Darmstadt. To this end, at least a rough 
binning of the triple-differential dilepton production cross section is 
needed. This requires high statistics, which is not easily achieved for such a rare 
probe. On the other hand, the angular distributions provide
valuable additional information, which can help disentangle the various 
contributions to the dilepton production cross section and thus also provide novel 
information on the properties of baryon resonances.

Several aspects 
of our model need to be improved in the future. First of all the model dependence 
of the predictions needs to be addressed. Repeating the 
calculation with a different choice for the $\rho$-baryon interaction Lagrangians, or a complementary approach, formulated in terms of helicity amplitudes or partial wave amplitudes, could provide
a more systematic framework for exploring the various contributions to the scattering amplitude.

A previous study suggests that at the CM energy of the HADES experiment a significant
part of the pion photoproduction cross section may be due to non-resonant Born contributions 
\cite{Zetenyi:2012hg}. Consequently, these Born terms can influence 
also the angular distributions of dilepton production and the anisotropy coefficient
in pion-nucleon collisions. Thus, their contribution to $\lambda_\theta$ should
be assessed.

Additional constraints on the model could be provided by studying one-pion and two-pion production in pion-nucleon
collisions, which have been measured at HADES with much better
statistics than the dilepton final state.  

\section*{Acknowledgments}
The work of E.S. was supported by VH-NG-823, Helmholtz Alliance HA216/EMMI and GSI. M.Z. was supported by the Hungarian OTKA Fund No. K109462 and EMMI. The work of B.F. was partially supported by EMMI.

\end{document}